\documentclass[fleqn,usenatbib]{mnras}

\usepackage{newtxtext,newtxmath}
\usepackage[T1]{fontenc}
\usepackage{ae,aecompl}

\newcommand{\elenct}{Lenc et al. (2018)\,}

\usepackage{graphicx}	
\usepackage{amsmath}	
\usepackage{amssymb}	
\usepackage[figuresright]{rotating}
\usepackage{caption}
\usepackage{hyperref}

\title[Radio detectability of exoplanets]{The detectability of radio emission from exoplanets}

\author[C. R. Lynch et al.]{
C. R. Lynch,$^{1, 2}$\thanks{E-mail: clynch@physics.usyd.edu.au}
Tara Murphy$^{1, 2}$
E. Lenc,$^{1, 2}$
D. L. Kaplan$^{3}$
\\
$^1$ Sydney Institute for Astronomy, School of Physics, The University of Sydney, NSW 2006, Australia\\
$^2$ ARC Centre of Excellence for All-sky Astrophysics (CAASTRO)\\
$^3$ Department of Physics, University of Wisconsin--Milwaukee, Milwaukee, WI 53201, USA\\
}

\date{Accepted 2018 April 27. Received 2018 April 25; in original form 2018 March 13}

\pubyear{2018}

\begin{document}
\label{firstpage}
\pagerange{\pageref{firstpage}--\pageref{lastpage}}
\maketitle

\begin{abstract}
Like the magnetised planets in our Solar System, magnetised exoplanets should emit strongly at radio wavelengths. Radio emission directly traces the planetary magnetic fields and radio detections can place constraints on the physical parameters of these features. Large comparative studies of predicted radio emission characteristics for the known population of exoplanets help to identify what physical parameters could be key for producing bright, observable radio emission. Since the last comparative study, many thousands of exoplanets have been discovered. We report new estimates for the radio flux densities and maximum emission frequencies for the current population of known exoplanets orbiting pre-main sequence and main-sequence stars with spectral types F-M. The set of exoplanets predicted to produce observable radio emission are Hot Jupiters orbiting young stars. The youth of these system predicts strong stellar magnetic fields and/or dense winds, which are key for producing bright, observable radio emission. We use a new all-sky circular polarisation Murchison Widefield Array survey to place sensitive limits on 200\,MHz emission from exoplanets, with $3\sigma$ values ranging from 4.0\,--\,45.0\,mJy. Using a targeted Giant Metre Wave Radio Telescope observing campaign, we also report a $3\sigma$ upper limit of 4.5\,mJy on the radio emission from V830 Tau b, the first Hot Jupiter to be discovered orbiting a pre-main sequence star. Our limit is the first to be reported for the low-frequency radio emission from this source.   

\end{abstract}




\section{Introduction}
Over the past few decades there has been a great effort to discover planets outside our Solar System, with more than 3700 systems now identified \citep{Schneider:2011}. The large contrast in optical or infrared brightness between the exoplanets and their host stars make discoveries through direct imaging difficult. Thus the majority of the currently known exoplanet population were discovered indirectly, through searches for the influence of the exoplanet on its host star \citep{Perryman:2011}. An alternative method of direct detection of magnetised exoplanets is through radio observations, as the expected radio emission produced by these planets could exceed the emission of the host star \citep{Griessmeier:2005}. Radio observations would also provide a direct measurement of the planet's surface magnetic field strength and in turn provide insight into the interior composition of these planets. Further, \citet{Hess:2011} show that the variability of the radio emission in both time and frequency can provide constraints on the rotational period of the planet, the orbital period and inclination, and the magnetic field tilt relative to the rotation axis.

The magnetised planets in our own Solar System are observed to emit intense, low-frequency radio emission from their auroral regions through the electron--cyclotron maser instability (CMI). The observed emission is highly circularly or elliptically polarised, beamed, and variable on time--scales ranging from seconds to days \citep{Wu:1979, Treumann:2006}. This emission arises from the propagation of energetic (keV) electrons along converging magnetic field lines in the planet's magnetosphere. Similarly, magnetised exoplanets are expected to emit intense, low-frequency radio emission \citep[e.g.][]{Winglee:1986, Zarka:2001}. 

The best targets for radio observation can be selected through theoretical estimates of the main characteristics of their planetary radio emission. The empirical relation called the Radiometric Bode$'$s law (RBL), based on observations of the magnetised Solar System planets, is primarily used to predict the intensity of radio emission from exoplanet. This relation relates the incident energy flux of the stellar wind to the radio power produced by a planet \citep{Desch:1984, Zarka:2001}.  However, the connection between the stellar wind and planetary auroral radio emission may not be as direct in large, co-rotating magnetospheres as it is for Jupiter. Jovian aurorae are driven by currents that form in the middle magnetosphere \citep{Nichols:2011}, although there is some evidence that the Solar wind may have an indirect effect on the produced Jovian emission \citep{Gurnett:2002}.

There have been a number of theoretical studies that estimate the expected radio flux densities from exoplanet systems using the RBL. Focusing on five massive planets orbiting Solar-like stars, \citet{Farrell:1999} noted $\tau$ Bootes to be an optimal target. \citet{Lazio:2004} modelled the radio flux density for 118 sources (known exoplanet population as of 2003 July 01), finding planets with small orbital distances to produce mJy level emission at frequencies between 10 $--$ 1000\,MHz.  Work by \citet{Stevens:2005} and \citet{Griessmeier:2005} focused on investigating how variations in the stellar wind and mass-loss rate affect the expected radio flux density. In the case of kinetic energy driven RBL, they find that objects with higher mass--loss rates and wind velocities relative to the Sun are more favourable targets for radio detections. This result suggests that Hot Jupiters located in more exotic planetary environments, such as around pre-main sequence stars \citep{Vidotto:2010,Vidotto:2017} or stars that have evolved off the main-sequence \citep{Fujii:2016} may be ideal candidates for radio detections. 

\citet{Griessmeier:2007a} compared the predicted radio flux densities for 197 exoplanets (known population of exoplanets as of 2007 January 13) using models where the level of planetary radio emission is related to the incident magnetic flux, incident kinetic energy of the stellar wind, or the incident energy from coronal mass ejections. They found that these different models lead to very different results, with the magnetic energy model predicting the largest flux densities.  They note, however, that all energy input models should be considered since it is not clear which dominates in planet--star systems. \citet{Griessmeier:2011} extended this analysis by predicting the flux densities for an updated catalogue of 547 objects (known population of exoplanets as of 2011 April 28). 

\citet{Nichols:2011, Nichols:2012} suggest that radio emission from some exoplanets may be dominated by a magnetosphere--ionosphere coupling current system associated with an internal plasma source such as an active moon. The expected radio emission from these planets would then not follow the RBL. They find that in such systems, fast rotating massive planets orbiting in large orbits around stars, with bright emission at X-ray to far-UV wavelengths, are capable of generating detectable radio emission. Further arguments against the RBL suggest that for close-in planets magnetospheric convection will saturate. These systems will not be able to dissipate the total incident stellar wind energy and thus the RBL will over estimate the flux densities substantially in these cases \citep{Jardine:2008, Nichols:2016}.

In addition to these theoretical studies, there have been many observational attempts to detect radio emission from exoplanets. The first searches for radio emission from exoplanets occurred before the first detection of the currently known population of exoplanets \citep{Winglee:1986}. Many of the more recent searches have involved targeting nearby Hot Jupiters previously detected through radial velocity and transit observations \citep{Bastian:2000, George:2007, Smith:2009, Lazio:2010, Stroe:2012, Lecavelier:2013, Hallinan:2013, Lynch:2017, OGorman:2018}. Others have used observations from low-frequency sky surveys to search for emission at the location of known exoplanets \citep{Lazio:2004, Sirothia:2014, Murphy:2015}. Despite these efforts there have been no unambiguous detections to date.

In this paper we update previous attempts to predict the radio emission from exoplanets by applying the RBL to the current population of known exoplanets. For systems with predicted emission properties that make them ideal candidates for detection, we place limits on their radio emission at 200 MHz using a recent Murchison Widefield Array (MWA; \citealt{Tingay:2013}) all-sky circular polarisation survey \citep{Lenc:2018}. We also report the results of a deep Giant Metrewave Radio Telescope (GMRT) observing campaign of a particularly interesting target, V830 Tau b, the first confirmed Hot Jupiter orbiting a pre-main sequence star \citep{Donati:2016}. 

\section{Radio emission from exoplanets}\label{sec:model}
As described by \citet{Zarka:2007}, there are in principle four different types of interaction between a planetary obstacle and an ambient stellar wind. These depend on whether the stellar wind and planet are magnetised or un-magnetised. In three of the four cases it is possible to produce intense non-thermal radio emission; the only case where radio emission is not produced is when an un-magnetised ambient wind encounters a un-magnetised planet. 

For cases where the star--planet interaction is expected to produce radio emission, the RBL suggests that the expected radio power, $P_{\text{radio}}$, is roughly proportional to the input power, $P_{\text{input}}$, from the ambient stellar wind
\begin{equation}
P_{\text{radio}}\ \propto P_{\text{input}}
\end{equation}
In the same manner as \citet{Griessmeier:2005,Griessmeier:2007a,Griessmeier:2007b}, we determine the proportionality constant by scaling the Jovian auroral radio emission, $P_{\text{radio,J}}$, with the input energy from the stellar wind. Observations of Jovian auroral emission show that the observed radio power varies greatly over time, with emission reaching powers as high as 4.5$\times$10$^{18}$\,erg s$^{-1}$. We use the average  power during periods of high activity as a reference value, with $P_{\text{radio,J}}$\,=\,2.1$\times$10$^{18}$\,erg s$^{-1}$. While scaling RBL using Jupiter's emission is standard practice, note that the majority of Jovian emission is not driven by an interaction with the Solar wind but by the planet's rotation \citep{Nichols:2011}. This limitation should be kept in mind throughout the rest of this work.

The star--planet interaction can be decomposed into either the dissipation of the kinetic power, $P_{\text{input,kin}}$, of the stellar wind protons impacting the planet's magnetosphere
\begin{equation}\label{eq:kin}
P_{\text{input,kin}} = m_pn v_{\text{eff}}^3\ \pi R_{m}^2
\end{equation}
or the interplanetary Poynting flux on the planet's magnetosphere
\begin{equation}\label{eq:mag}
P_{\text{input,mag}} = (B_{\perp}^2/8\pi)\ v_{\text{eff}}\ \pi R_{m}^2.
\end{equation}
Here $m_p$ is the mass of a proton, $n$ the number density of the ambient stellar wind at the location of the planet, $v_{\text{eff}}$ the effective velocity of the stellar wind in the reference frame of the planet, $R_m$ is the radius of the planetary magnetosphere, and $B_{\perp}$ is the interplanetary magnetic field perpendicular to the stellar wind flow. The radio flux density, $S_{\nu}$, expected from a magnetised exoplanet at a distance $d$ is then
\begin{equation}
S_{\nu} = \frac{P_{\text{radio}}}{\Omega d^2 \Delta\nu}
\end{equation}
where $\Omega$ is the solid angle of the beam of emitted radiation, which we assume to be equivalent to that of Jovian radio emission $\Omega$=1.6 sr \citep{Zarka:2004} and $\Delta\nu$ is the bandwidth of the emission and is taken to be the maximum emission frequency. Depending on the type of star--planet interaction, the radio power will be given by scaling equation (\ref{eq:kin}) or (\ref{eq:mag}) by $P_{\text{radio,J}}$. 

Planetary CMI emission is produced near the local cyclotron frequency within the source region of the magnetosphere. The CMI radio spectrum is expected to show a sharp cutoff at the cyclotron frequency associated with the maximum magnetic field strength, $B_p^{\text{max}}$, close to the planet's surface. The maximum emission frequency, $\nu_c^{\text{max}}$, is then
\begin{equation}
\nu_c^{\text{max}} = \frac{eB_p^{\text{max}}}{2\pi m_e}\approx 2.8\ \text{MHz}\ B_p^{\text{max}}.
\end{equation}
Here, $m_e$ and $e$ are the electron mass and charge, and $B_p^{\text{max}}$ is measured in Gauss.

For an observer to detect the radio emission from an exoplanet, in addition to being bright enough, the emission must also be able to propagate from the source to the observer. The characteristic plasma frequency, $\nu_p$, of a plasma with number density $n$ (in cm$^{-3}$),
\begin{equation}
\nu_{p} = \sqrt{\frac{ n e^2}{\pi m_e}} \approx 8.98\ \text{KHz}\ \sqrt{n}
\end{equation}
provides a lower frequency limit for the propagation of electromagnetic radiation. Emission at frequencies lower than the local plasma frequency will be absorbed by the plasma. In order for the emission to propagate from the exoplanet system, the local cyclotron frequency within the source region, $f_c^{\text{max}}$, must be greater than the local plasma frequency at every point along the line of sight
\begin{equation}\label{eq:compare}
\nu_{p}^{\text{max}} < \nu_c^{\text{max}}.
\end{equation}
Because the electron density of the stellar wind decreases with distance, this condition is most restrictive at the orbital distance of the planet. Thus to determine if CMI emission is able to reach distant observers, it is sufficient to check that equation (\ref{eq:compare}) is satisfied at the orbital location of the exoplanet. The emission frequency must also be larger than the plasma frequency of Earth's ionosphere for it to be observable by ground based telescopes. Earth's ionosphere has a number density $<$10$^6$cm$^{-3}$ corresponding to a maximum plasma frequency of 10\,MHz. 

An additional factor that affects whether CMI emission is generated, is the ratio of the plasma frequency to the electron cyclotron frequency at the location of the radio wave generation. CMI is only efficient at generating intense radio emission in density-depleted regions or regions of high magnetic field strength. In the preferred source regions, the in situ plasma frequency is small in comparison to the local electron cyclotron frequency \citep{Treumann:2006}. The exact ratio required is still uncertain; observations favour a ratio of $\nu_p$/$\nu_c$$\lesssim$0.1, while CMI theory supports a ratio closer to 0.4 \citep{LeQueau:1985, Hilgers:1992, Zarka:2001}. We assume that at some location along the poloidal magnetic field line, the ratio of the plasma to cyclotron frequency is such that the CMI mechanism is able to efficiently produce radio emission. 

\section{Model parameters}

To estimate the expected radio flux densities and emission frequencies of exoplanet emission, we need to describe the properties of the stars, including the stellar wind properties, and the properties of the exoplanets. We take the host star mass, $M_*$, age, $t$, and distance from Earth, $d$, from the Extrasolar Encyclopaedia \citep{Schneider:2011} or the NASA Exoplanet Archive\footnote{\url{https://exoplanetarchive.ipac.caltech.edu}}. In most cases we can also take the star's radius, $R_*$, directly from these catalogues. For main-sequence stars without a listed radius we use the listed effective stellar temperature and mass to estimate $R_*$, assuming a mass-luminosity relation of $L\propto M^4$ \citep{Duric:2004}; pre-main sequence stars without measured radii are removed from our sample.   

For most of the exoplanets the observed mass listed is the projected mass of the planet, $M_{\text{obs}}$\,=\,$M_p \sin(i)$, where $i$ is the inclination angle of the planet's orbit with respect to the observer. The distribution function of inclinations for a randomly distributed sample, ignoring selection effects, is proportional to $\sin(i)$ \citep{Halbwach:1987}. For exoplanets with only a projected masses, we use the expected value, $<\sin i>$, to estimate the real mass of the exoplanets, $M_p$\,=\,$M_{\text{obs}}$\,$\cdot$\,$(\frac{1}{<\sin i>})$\,$\approx$\,1.15\,$M_{\text{obs}}$; otherwise we use the observed mass as the mass of the planet, $M_p$. 

The expected radio flux density for exoplanets with large eccentricities can vary by factors of 2\,--\,3 over the course of the planet's orbit \citep{Stevens:2005}. Thus we use the eccentricity, $e$, and semi-major axis, $a$, listed in the on-line catalogues for each exoplanet, to calculate the periastron distance, $a_{\text{min}}$\,=\,$\frac{a}{(1-e)}$. The estimated radio flux densities reported here are only for this orbital distance. 

The planetary radius is required for estimating both the planetary magnetic field strength and flux density of the expected radio emission. For transiting planets, we can use the measured radius, $R_p$, listed in the online exoplanet catalogues. For non-transiting planets we calculate the planetary radius following the method outlined in \citet{Griessmeier:2007a}. Using the results of numerical simulations for planetary radii, \citet{Griessmeier:2007a} derived analytical expressions to calculate the radius of a planet of mass $M_p$, with an orbital semi-major axis $a$. For a `cold' planet
\begin{equation}
\begin{split}
R_p(a = \infty) &= \frac{\left(\alpha M_p\right)^{1/3}}{1 + \left(\frac{M_p}{(M_{max}/M_J)}\right)^{2/3}} \\ 
&\approx 1.47 R_J \frac{(M_p/M_J)^{1/3}}{1 + \left(\frac{(M_p/M_J)}{(M_{\text{max}}/M_J)}\right)^{2/3}}
\end{split}
\end{equation}
where Jupiter's radius is taken to be $R_J$\,=\,7.1492$\times 10^9$\,cm, $\alpha$\,=\,0.61\,cm$^3$\,g$^{-1}$ for a planet with the same composition as Jupiter, $M_{\text{max}}$\,=\,3.16\,$M_J$, and we use $M_J$\,=\,1.9$\times 10^{30}$\,g. The radius of an irradiated planet is then given by
\begin{equation}
\frac{R_p(a)}{R_p(a=\infty)} = \frac{(R_p(a)/R_J)}{(R_p(a=\infty)/R_J)} = \left[1 + 0.05\left(\frac{T_{eq}}{T_0}\right)^{\gamma}\right].
\end{equation}
Here $T_{eq}$ is the equilibrium temperature of the planet's surface and the coefficients $T_0$ and $\gamma$ depend on the planetary mass (see appendix A of \citet{Griessmeier:2007a} for more details).

\subsection{Stellar Wind Model} 
The stellar wind velocity and density encountered by an orbiting magnetised exoplanet plays a significant role in determining both the size of the planet's magnetosphere, as discussed in section (\ref{sec:pmag}), and the amount of energy available for the generation of the planetary radio emission. Close to the star, the stellar wind is still accelerating and provides much less input energy for the generation of planetary radio emission than assuming the quasi-asymptotic wind velocity of the Sun \citep{Griessmeier:2007a}. To avoid overestimating the expected radio flux densities for small orbital distances, as in the cases of Hot Jupiters, it is necessary to use a distance dependent stellar wind model. For simplicity we employ Parker's isothermal solution for the stellar wind speed, $v(r)$, at a radial distance $r$ from the surface of the star \citep{Parker:1958}
\begin{equation}\label{eq:parker}
\frac{v(r)^2}{c_s^2}-\ln\left(\frac{v(r)^2}{c_s^2}\right) = 4\ln\left(\frac{r}{r_c}\right)+4\frac{r_c}{r} -3. \end{equation}
The wind velocity will pass through sound speed $c_s$, given by
\begin{equation}
c_s = \sqrt{\frac{k_bT}{m_p}}
\end{equation}
at the critical distance $r_c$
\begin{equation}
r_c = \frac{m_pGM_*}{4k_bT}
\end{equation}
where $k_b$ is the Boltzmann constant, $G$ is the gravitational constant, $T$ is the temperature of the stellar wind and equivalent to the coronal temperature for our isothermal model, and $M_*$ the mass of the star. In this work we assume the stellar wind to be composed of protons.

\citet{Griessmeier:2007a} showed that the Parker stellar wind model is sufficient for describing the radial properties of winds for stars with ages $>$0.7\,Gyr. The youngest objects in our sample are much younger than 0.7\,Gyr, including 8 non-accreting pre-main sequence stars (i.e. weak line T Tauri (WTT) stars). To check the radial velocity profiles for the young stars in our sample we compared them to the profiles resulting from three-dimensional numerical magnetohydrodynamic simulations of WTT stellar winds by \citet{Vidotto:2010}. Using the case of a magnetic field aligned with the rotation axis, we estimate a difference in the expected velocities of $\sim$10$\%$ at distances $<$10$R_*$ and up to a factor of 2 at large distances ($>$60$R_S$); this is sufficient for our order-of-magnitude estimates of the planetary radio emission flux densities.

To determine the stellar wind velocity at a distance $r$, we solve equation (\ref{eq:parker}) numerically, where the coronal temperature is a free parameter. To determine the coronal temperature for main-sequence stars we follow the method outlined in \citet{Griessmeier:2007a}: First we determine the stellar wind velocity expected at 1\,AU given the star's age and using the expression for the time behaviour of the Solar wind at 1\,AU \citep{Newkirk:1980},
\begin{equation}\label{eq:newkirk}
v(1\ \text{AU},t) = v_0\left(1 +\frac{t}{\tau}\right)^{-0.43}
\end{equation}
whose proportionality constants, $v_0$\,=\,3.971$\times 10^8$\,cm\,s$^{-1}$ and $\tau$\,=\,2.56$\times$10$^7$\,yr, are determined by present-day Solar conditions. Next we adjust the coronal temperature until the corresponding stellar wind velocity at 1\,AU found using Parker's solution (equation \ref{eq:parker}) agrees with the value we calculate from equation (\ref{eq:newkirk}). We then use this temperature to solve for the final radial profile of the stellar wind velocity for a range of distances between 0.09 and 200 times the critical distance $r_c$. For pre-main sequence stars we assume the coronal temperature to be 10$^6$\,K in agreement with the models of \citet{Vidotto:2010} and solve for the stellar wind radial profile over the same distance range. Each wind profile is plotted and visually inspected to ensure that the physically meaningful solution of Parker's wind model is found. 

For planets orbiting at small distances from their host stars, the Keplerian velocity of the planet will be comparable to the stellar wind velocity. These orbits are expected to be circular due to tidal dissipation and the Keplerian velocity is then given by
\begin{equation}
v_{\text{k}}(r) = \sqrt{\frac{M_* G}{r}}.
\end{equation}
The effective velocity of the stellar wind plasma relative to the motion of the planet is then
\begin{equation}
v_{\text{eff}} = \sqrt{v_{\text{k}}(r)^2 + v(r)^2}.
\end{equation}

The radial distance dependence of the stellar wind density is determined using the mass-loss rate, $\dot{M}_*$, expected for each star
\begin{equation}
n(r) = \frac{\dot{M}_*}{4\pi m_pv(r)r^2}
\end{equation}
where $v(r)$ is the stellar wind velocity at the orbital distance of the exoplanet. 

The stellar mass-loss rate is found independent of the wind solution and is based on the age of the star. Charge exchange interactions between the neutral hydrogen of the local interstellar medium and an ionised stellar wind creates a population of hot neutral hydrogen surrounding the star. The absorption of this hot neutral hydrogen can be detected in high-resolution Lyman--$\alpha$ spectra. The amount of absorption measured can be used as a diagnostic for the stellar mass-loss rate \citep{Wood:2002}. Using Solar-type stars with both measured X-ray luminosities and HI Lyman--$\alpha$ absorption, \citet{Wood:2002, Wood:2005} observed an empirical relationship between the mass-loss rate of these stars and their coronal X-ray surface flux. Because the coronal X--ray surface flux is observed to evolve with stellar age, \citet{Wood:2002, Wood:2005} then related the stellar mass-loss rates to their age, $\dot{M}$\,$\propto$\,$t^{2.33\pm0.55}$. This scaling relationship is only valid for stars with X--ray fluxes $F_X$\,$<$\,10$^6$\,erg\,cm$^{-2}$\,s$^{-1}$, and excludes the youngest and most active Solar-type stars \citep{Wood:2002, Wood:2005}. 

Alternatively, \citet{Alvarado:2016a, Alvarado:2016b} combined Zeeman--Doppler imaging of magnetic structures for young Solar-type stars and numerical codes used for Solar system space weather modelling, to develop a self-consistent and data-driven characterisation of the wind environment for young active Solar-type stars. These models were then used to derive a fully-simulated, mass-loss activity relation 
\begin{equation}
\dot{M}\propto F_X^{0.79\substack{+0.19 \\ -0.14}}.
\end{equation}.
Using the relation between the X--ray surface flux and age, $F_X$\,$\propto$\,$t^{1.74\pm0.34}$ \citep{Ayres:1997}, this results in the following scaling relationship for the mass-loss evolution of main-sequence Solar-type stars:
\begin{equation}\label{eq:mdot}
\dot{M}_*\propto t^{-1.37}.
\end{equation}
This expression is valid for high-activity stars with X--ray fluxes up to 100 times that of the Sun. Scaling equation (\ref{eq:mdot}) using the measured mass-loss rate for the Sun, $\dot{M}_{\odot}$\,=\,10$^{14}$\,$M_{\odot}$\,yr$^{-1}$, we use it to calculate the mass-loss rates for the main-sequence stars in our sample. For the WTT stars in our sample we assume the mass-loss rate to be 10$^{-9}$\,$M_{\odot}$\,yr$^{-1}$, consistent with the modelling of winds from WTT stars by \citet{Vidotto:2010}.

For a Parker spiral, the interplanetary magnetic field, ($B_r$,$B_{\phi}$), at a distance $r$ from the surface of the star is given by
\begin{equation}
B_r = B_0\ \left(\frac{r_0}{r}\right)^2
\end{equation}
and
\begin{equation}
B_{\phi} = B_r\ \frac{\Omega r}{v_{\text{eff}}}
\end{equation}
where $r_0$\,=\,$R_*$, $B_0$ is the magnetic field strength at the surface of the star, and $\Omega_s$\,=\,2$\pi$/$P_s$ is the angular velocity of a star with rotational period $P_s$. The interplanetary magnetic field perpendicular to the stellar wind flow can then be calculated using
\begin{equation}
B_{\perp} = \sqrt{B_r^2 + B_{\phi}^2}\ \sin\left[\arctan\left(\frac{B_{\phi}}{B_r}\right) - \arctan\left(\frac{v_{\text{k}}}{v}\right)\right].
\end{equation} 

main-sequence Solar-type stars are thought to lose significant angular momentum through the coupling of their magnetic fields and wind, causing them to spin down over time \citep{Bouvier:2014}. Similar to \citet{Griessmeier:2007b}, we estimate the rotational periods for the main-sequence stars using
\begin{equation}
P_s \propto \left(1 + \frac{t}{\tau}\right)^{0.7}
\end{equation}
where $t$ is the system age, and we take the time constant $\tau$\,=\,2.56$\times$10$^7$\,yr and the rotational period of the Sun $P_{\odot}$\,=\,25.5\,days. 

The rotational evolution of young, still forming stars is quite different from the scenario for the older main-sequence stars. In the first few Myr of a pre-main sequence star's evolution, its rotation is determined by interactions between the forming star and its circumstellar disk, which removes angular momentum as the star contracts towards the main-sequence. The circumstellar disks are expected to dissipate on a timescale of a few Myr \citep[e.g.][]{Haisch:2000} and the rotation of the pre-main sequence star is predicted to become faster as the young star continues its contraction but no longer has an efficient braking mechanism \citep{Karim:2016}. The rotational periods of pre-main sequence stars have been measured in several star forming regions, showing a large scatter in the rotational properties among stars that are part of a supposedly coeval population. This scatter could be the result of a wide range of initial conditions, like the primordial disk fraction, and hint at the importance of environmental effects on rotation properties \citep{Karim:2016}. Of the 8 WTT stars in our sample only V830 Tau b has a previously measured rotation period of 2.741 days \citep{Donati:2016}. For the other 7 objects we use a rotational period of 4.31 days, which is the average rotational period measured for the WTT stars found in the Taurus--Auriga star forming region \citep{Grankin:2016}.

Using literature magnetic field results, \citet{Vidotto:2014} investigated how the large-scale surface magnetic fields of low-mass stars ($\sim$0.1 - 2\,$M_{\odot}$), reconstructed via Zeeman--Doppler imaging, vary with age. The sample \citet{Vidotto:2014} used consisted of 73 stars ranging in evolutionary stages from accreting pre-main sequence stars to main-sequence objects (1\,Myr\,$<$\,$t$\,$<$10\,Gyr). For the non-accreting dwarf stars they find that the unsigned average large-scale surface field scales with age $t$ as
\begin{equation}\label{eq:surfB}
<|B_v|>\propto t^{0.655\pm0.045}.
\end{equation}
This correlation holds over two orders of magnitude in $<|B_v|>$ and three orders of magnitude in $t$. \citet{Folsom:2016} did a similar study which focused on a more well-defined sample of stars, with ages established from clusters or co-moving groups, finding a correlation between magnetic field strength and age that agrees with \citet{Vidotto:2014}. To estimate the surface magnetic field strengths, $B_0$, for the host stars in our sample we scale equation (\ref{eq:surfB}) using the appropriate values for the Sun's age, $t_{\odot}$\,=\,4.6\,Gyr, and minimum large-scale surface magnetic field, 
$<|B_{v,\odot}|>$\,=\,1.89\,G.

\subsection{Planetary magnetosphere}\label{sec:pmag}

Predictions of the maximum emission frequency for the planetary radio emission require estimates of the planet's magnetic field strength at its surface. Scaling relations are used to relate the magnetic field intensity of a planet's dynamo to the physical properties of the planet. \citet{Griessmeier:2007a,Griessmeier:2007b,Griessmeier:2011} use scaling relations that assume a balance of Coriolis force and Lorentz force; in these relations the magnetic field strength depends on the rotation rate of the planet. 

In contrast, \citet{Christensen:2006} assume that the magnetic field strength is determined by the energy flux through the dynamo region and is independent of the planetary rotation rate. Using direct numerical simulations of planetary dynamos over a wide range of parameters, \citet{Christensen:2010} tested the predicted magnetic field strengths of various scaling relations. These tests showed that magnetic field scaling based on the available energy flux better matched the results of the numerical simulations. Further \citet{Christensen:2009} generalised the energy flux scaling relation and showed that the predicted magnetic field strengths agree with observations of a wide range of rapidly rotating objects, from Earth and Jupiter to brown dwarfs and low-mass main-sequence stars. In the following, we choose to use the scaling relation of \citet{Christensen:2009} to predict the magnetic field strengths for our sample of exoplanets. 

For our sample of exoplanets we use the simple expression provided by \citet{Reiners:2010} that relates the mean magnetic field strength at the surface of the dynamo to a planet's mass, radius, and luminosity $L_p$
\begin{equation}
B_{\text{dyn}} = 4.8\ \text{kG}\ \left[\frac{(M_p/M_J) (L_p/L_{\odot})}{(R_P/R_J)^7}\right]^{1/6}.
\end{equation}
For massive brown dwarfs and stars, the dynamo surface is located close to the object's surface and $B_{\text{dyn}}$ represents the mean surface magnetic field. However for giant planets the surface of the dynamo region lies within the interior of the planet and the surface magnetic field is attenuated due to the overlaying material. \citet{Reiners:2010} assume that the depth of the dynamo surface is inversely proportional to $M_p$ and so the dipole field strength at the equator of the planet is related to the mean magnetic field strength at the surface of the dynamo by
\begin{equation}
B_{\text{dip}}^{\text{eq}}=\frac{B_{\text{dyn}}}{2\sqrt{2}}\left(1 - \frac{0.17}{M_p/M_J}\right)^3.
\end{equation}
The polar dipole field strength is then, $B_{p}^{\text{max}}\,=\,2\, B_{\text{dip}}^{\text{eq}}$.
The luminosity of each planet is calculated using the power-law expressions in \citet{Burrows:2001} for massive planets. 

The radius of the planetary magnetosphere is determined by the pressure balance between the planet's magnetic field and the external stellar wind pressure, including the wind thermal, magnetic, and ram pressure. 
\begin{equation}\label{eq:rmag}
\frac{B_p(r)^2}{8\pi} = 2nk_BT + \frac{B_*^2}{8\pi} + m_pnv_{\text{eff}}^2
\end{equation}
where $B_*$\,=\,$\sqrt{B_r^2\,+\,B_{\phi}^2}$ is the magnitude of the stellar magnetic field at the location of the planet. Assuming the planetary magnetic field is a dipole, the magnetic field strength, $B_p(r)$, scales with distance as $(R_p/R)^{3}$. Substituting this distance scaling and solving for the magnetospheric radius, we find
\begin{equation}\label{eq:magRad}
\frac{R_{m}}{R_p} = (2.44)^{1/3}\left[\frac{(B_{p}^{\text{max}})^2}{8\pi\left(m_pnv_{\text{veff}}^2+2nk_BT+\frac{B_*^2}{8\pi}\right)}\right]^{1/6}.
\end{equation} Included in equation (\ref{eq:magRad}) is a correction factor of 2.44, which accounts for the enhancement of the magnetic field at the magnetosphere boundary due to currents \citep{Mead:1964}. 

The magnetosphere of planets with weak magnetic field strengths or those embedded in dense stellar winds, as in the case for young systems, can be compressed to such an extent that its radius is less than the planet's radius. For these planets we assume the magnetospheric radius is equal to the planetary radius.

\begin{figure*}
\sbox0{\begin{tabular}{@{}cc@{}}
\includegraphics[width=115mm]{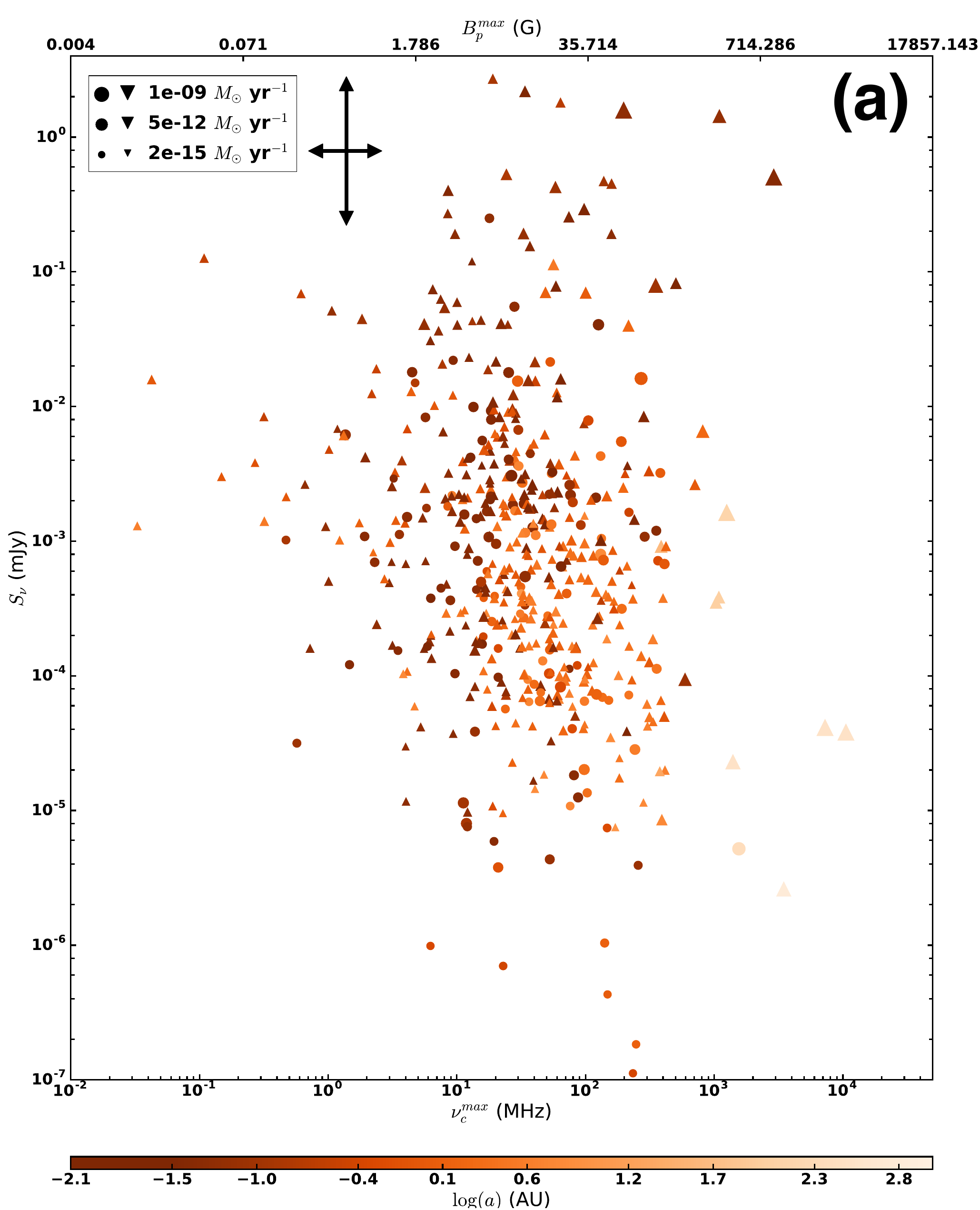}&
\includegraphics[width=115mm]{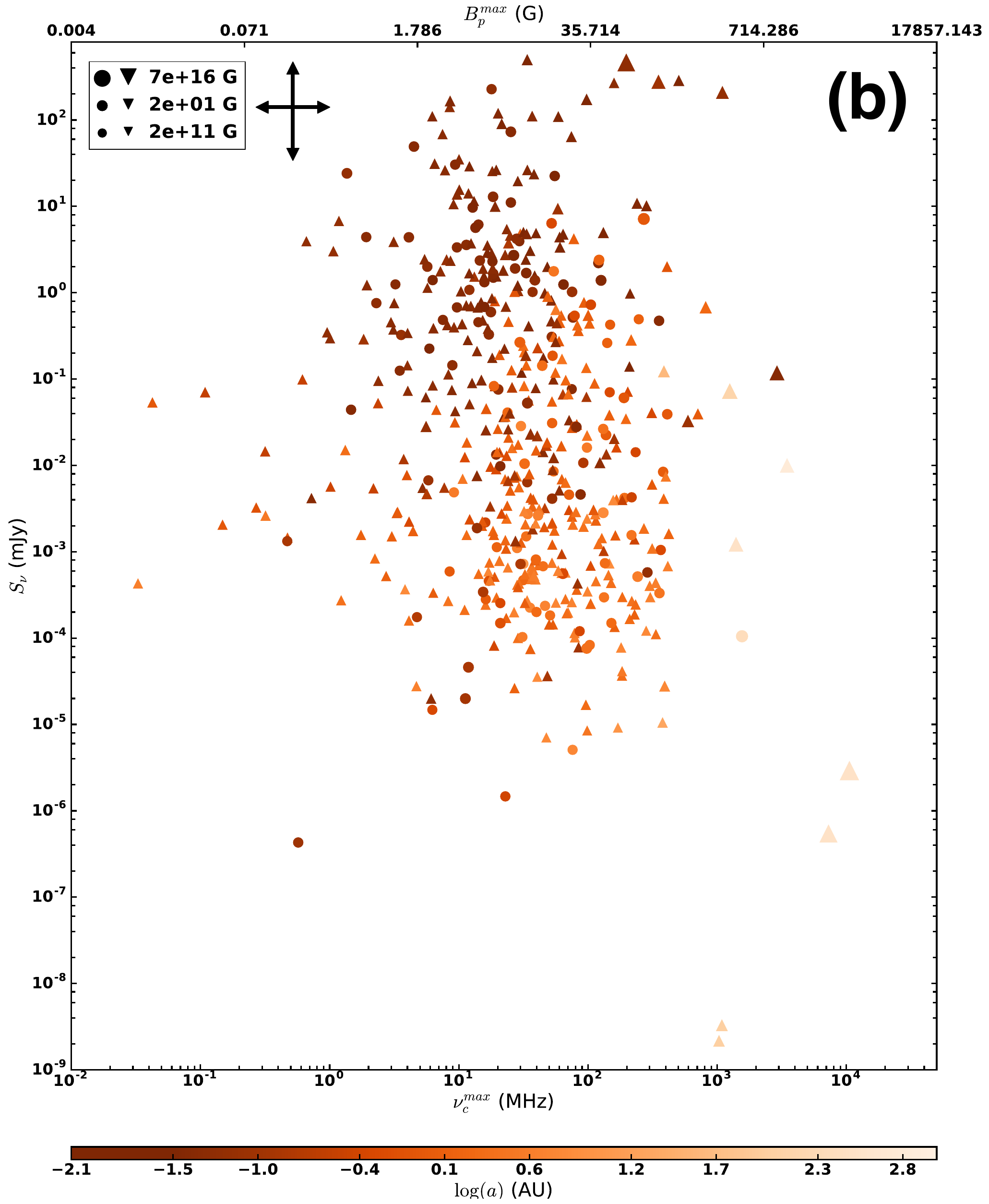}
\end{tabular}}
\rotatebox{90}{\begin{minipage}[c][\textwidth][c]{\wd0}
\usebox0 
\captionof{figure}{The predicted maximum observing frequency (bottom axis) and corresponding planetary magnetic field strength (top axis), and expected radio flux density using the kinetic (a) and magnetic (b) energy model for the current population of known exoplanets orbiting dwarf-type stars. The y-axis for each model is chosen to encompass the full sample and is different between these two figures. The size of each symbol corresponds to the predicted mass--loss rate (a) or surface magnetic field strength (b) of the host star; the colour corresponds to the natural log of the semi-major axis of the exoplanet system. Triangle symbols represent systems with declinations <+30$^{\circ}$, corresponding to the declination limit of the GLEAM survey (see section \ref{sec:mwa_obs}); exoplanets with circle symbols have declinations north of this limit. The typical uncertainties for these predicted values are given by the arrows in the top left corner.}\label{fig:full_pop}
\end{minipage}}
\end{figure*} 

\section{Exoplanet census}

To generate our sample of exoplanets we use the full catalogue of the Extrasolar Planet Encyclopaedia and the confirmed source catalogue for the NASA Exoplanet archive from 2018 February 19. After removing common sources we combine these two catalogues into a master sample of 4133 sources. Due to the stellar wind models we use to estimate the radio flux densities, we chose to focus on exoplanets orbiting main-sequence and pre-main sequence dwarf stars (spectral types F, G, K, and M) and remove all other exoplanet systems from our sample. After removing the non-dwarf type host stars we are left with 1287 sources. We also remove all exoplanets with masses greater than 13\,$M_J$ -- objects with masses greater than this are more likely to be brown dwarf objects than planets. Making this cut removes 35 objects. For terrestrial planets \citet{Driscoll:2011} modified the scaling relations from \citet{Christensen:2006} by assuming the planetary internal structure to be similar to that of the Earth and used an optimal mantle convection model that maximises the available heat flux to the planet's core to drive the dynamo. For this mass range, they found the planetary surface magnetic fields are reasonably constant up to planetary masses twenty times that of the Earth ($M_P$\,$<$\,0.06\,$M_J$) and roughly $\sim$3$B_{s,E}$, where $B_{s,E}$\,=\,0.3\,G. This magnetic field strength corresponds to maximum emission frequencies $<$\,10\,MHz and unobservable by ground-based radio telescopes. Thus we remove all sources with masses $M_P$\,$<$\,0.06\,$M_J$ from our sample; this removes 95 objects. 

There are several measured physical quantities that are required for calculating either the expected radio flux density or emission frequency, they include: both the mass of the exoplanet and the star, the distance to the exoplanet system, the semi-major axis of the exoplanet orbit, and the age of the star. Exoplanets that are missing any one of these parameters are removed from our sample of sources. Pre-main sequence stars are removed if they do not have a measured stellar radius. Main-sequence stars are removed if they do not have a measured stellar radius and temperature; main-sequence stars with only one of these quantities remain in our sample. After making these selections this leaves us with a total of 597 exoplanets in our sample. 

We calculate the expected radio flux density and emission frequency for each of the 597 exoplanets in our sample. Of the 597 objects, 69 ($\sim$12$\%$) do not satisfy the propagation condition in equation (\ref{eq:compare}). The results for the remaining sources are shown in Figure (\ref{fig:full_pop}) for both the kinetic (\ref{fig:full_pop}a) and magnetic (\ref{fig:full_pop}b) RBL. In this figure the estimated radio flux density is plotted as a function of the maximum emission frequency (bottom axis) or the maximum planetary magnetic field (top axis). The colour of each symbol corresponds to the natural log of the semi-major axis for the star--planet system, objects with declinations $<+$30$^{\circ}$ are represented by a triangle symbols and those with declination $>+$30$^{\circ}$ correspond to the circles. The size of each symbol represents the estimated mass-loss rate (\ref{fig:full_pop}a) or surface magnetic field (\ref{fig:full_pop}b) of the host star based on its age. 

Comparing the estimated flux densities for the kinetic and magnetic RBL in Figure (\ref{fig:full_pop}), its clear that in most cases the magnetic RBL predicts radio flux densities at least order of magnitude brighter than for the kinetic RBL. However, there are a few exoplanets where the opposite is true -- adding further support to consider both magnetic and kinematic RBL when determining best candidates for radio detections.  Similar to the modelling results of both \citet{Lazio:2004} and \citet{Griessmeier:2007b, Griessmeier:2011}, we find a general trend of decreasing flux density as the maximum emission frequency increases. Generally the higher emission frequencies are associated with larger planetary masses \citep{Reiners:2010}, and from the colour--mapping its clear that the most massive planets in our sample are orbiting at large distances from their host star, resulting in low radio flux densities. This trend reflects a lack of high mass planets in orbits with small semi-major axes. This deficit of massive planets at small orbital radii is a well known phenomena, explained by strong tidal interactions that lead to a rapid decay of the planetary orbital radius over a short period of time, with the planet eventually reaching the Roche limit of the host star and is effectively destroyed \citep{Patzold:2002, Jiang:2003}. The outliers to this trend are those objects that are the best candidates for radio observations.

The uncertainties associated with the estimated radio flux densities and emission frequencies are considerable. They are represented by the arrows in the upper left corner of Figure (\ref{fig:full_pop}). Similar to \citet{Griessmeier:2007a} we estimate the uncertainty for the emission frequency and radio flux density by considering the different sources of error and calculate the range of results for a select group of exoplanets from our sample. Using the uncertainties reported in the NASA Exoplanet Archive and the Extrasolar Planet Encyclopaedia we find that the largest uncertainties are for the exoplanet mass and stellar age. We choose four sources (V830 Tau b, TAP 26 b, WASP--38 b, and tau Boo b) which span a range of ages between 0.002\,--\,5\,Gyr and planetary masses between 0.7\,--\,4.97\,$M_J$. The uncertainty in the exoplanet masses reported by these two archives can be as much as a factor of 3 and for the ages up to a factor of 4. We compare the emission frequencies and estimated flux densities using the kinetic and magnetic RBL to the original values, after varying these two parameters by a factor of 3 for the planet mass and a factor of 4 for the stellar age. We find that the variation in the planet mass has little effect on the estimated flux densities, changing them by factors $<$2, but can change the estimated emission frequencies by up to a factor of 3. We find that the variation in the age greatly affects the estimated flux densities using the kinetic RBL for all the sources with variations up to a factor of 6, however only the magnetic RBL values for the younger two sources (V830 Tau b and TAP 26 b) are significantly affected by the age variation with a change of up to a factor of 7. To be conservative we assume the uncertainty in our flux estimates to be an order of magnitude and for our estimated emission frequencies a factor of 3; these are in agreement with the modelling results of \citet{Griessmeier:2007a}. Given these large uncertainties, the estimated values reported here are intended to be a guide for future observations and are not considered to be accurate.

\begin{figure*}
\sbox0{\begin{tabular}{@{}cc@{}}
\includegraphics[width=115mm]{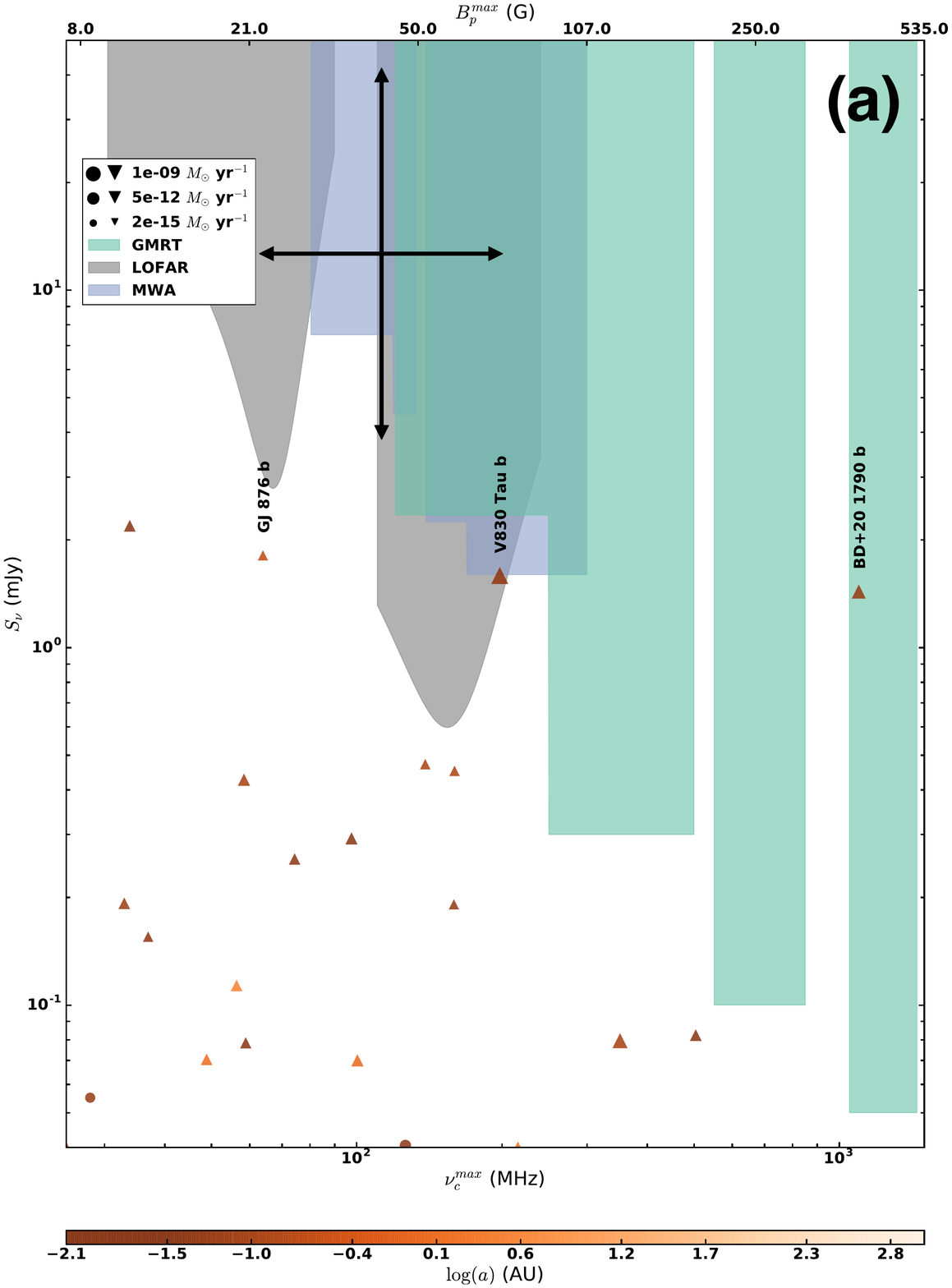}&
\includegraphics[width=114mm]{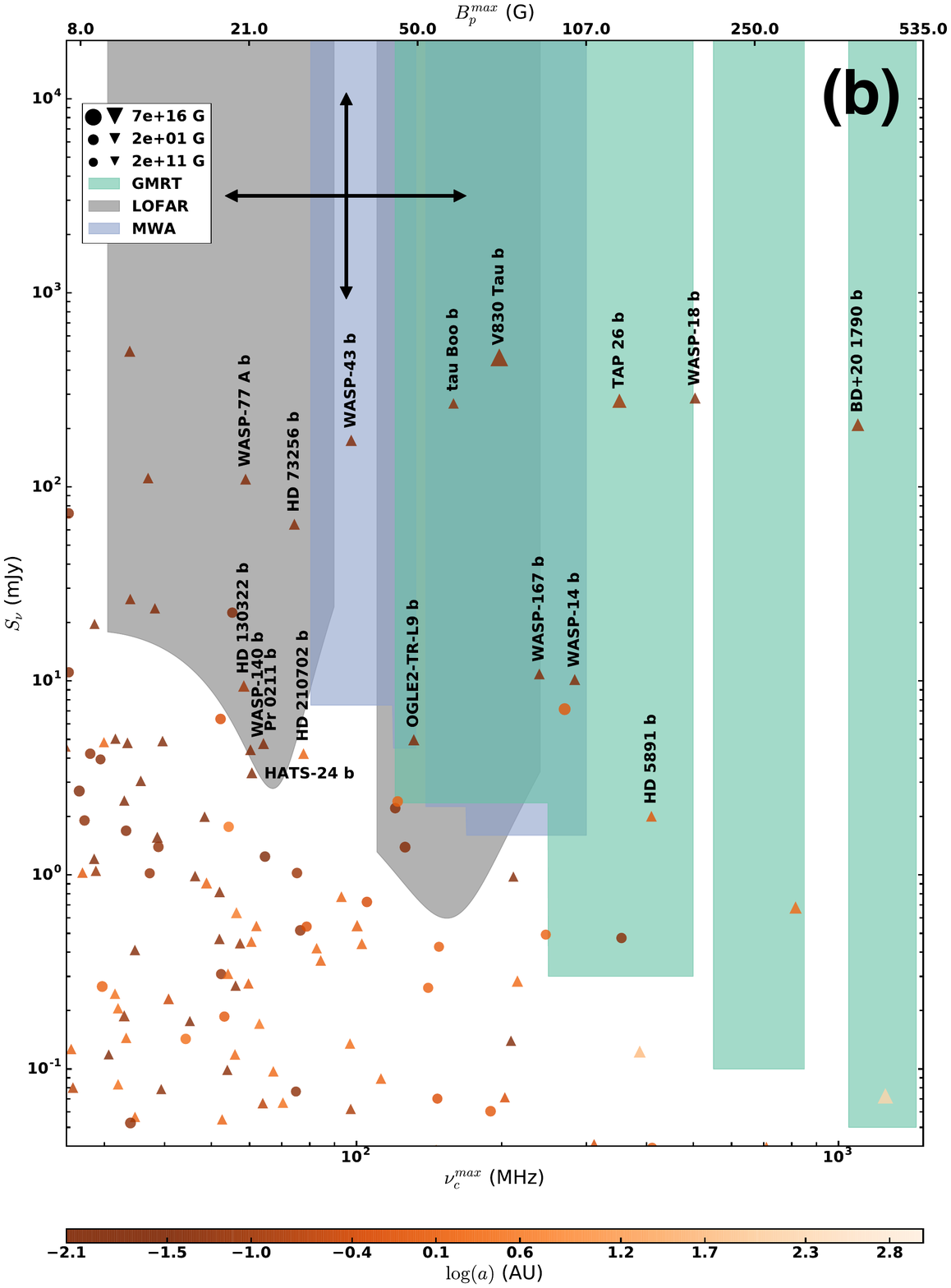}
\end{tabular}}
\rotatebox{90}{\begin{minipage}[c][\textwidth][c]{\wd0}
\usebox0
\captionof{figure}{The predicted maximum observing frequency (bottom axis) and corresponding planetary magnetic field strength (top axis), and expected radio flux density using the kinetic (a) and magnetic (b) energy model centred on objects expected to produce the brightest radio emission; again the range of values on the y-axis is different between the two figures. These sensitivities of three low-frequency radio telescopes are indicated by the green (GMRT), grey (LOFAR), and blue (MWA) shaded regions. The identified sources are those we analysed using the all-sky Stokes V MWA survey by \elenct. All other colours and symbols are as described in Figure (\ref{fig:full_pop}).}\label{fig:tele_compare}
\end{minipage}} 
\end{figure*}

Generally, the range of radio frequencies and expected radio flux densities shown in Figure (\ref{fig:full_pop}) agree with previous large comparative studies by \citet{Lazio:2004} and \citet{Griessmeier:2007b, Griessmeier:2011}. We see the same trends in the predicted radio flux densities as a function of the maximum emission frequency as these authors. However comparisons of individual sources reveal significant differences between our model and the models of these authors. Both \citet{Lazio:2004} and \citet{Griessmeier:2007b} use different scaling relations as to that used here to predict the maximum emission frequency leading to large difference between our results and theirs for specific sources. As previously discussed in the literature the scaling relation used by \citet{Lazio:2004} was disproven through observations of the Solar system and should not be used to predict planetary emission frequencies \citep{Griessmeier:2007b}. The scaling relation from \citet{Griessmeier:2007b} is rotationally dependent and generally predicts lower maximum emission frequencies than the rotationally independent relations we used here. The difference is not as large, and within the uncertainties of the predicted frequencies, if we consider the non-rotationally dependent frequencies predicted by \citet{Griessmeier:2011}, with differences of less than a factor of 3. To check our predicted emission frequencies we compared the results for the specific sources modelled by \citet{Reiners:2010}, finding our values to be within a factor of 2 of their predictions.

We use very different scaling relations for the mass-loss rate and stellar surface magnetic field as to that used by \citet{Griessmeier:2007b,Griessmeier:2011}. The relations chosen in this work are more appropriate for estimating these physical quantities over a larger range in stellar age, down to a few Myr, and allowed for more accurate predictions of the emission for the young sources in our sample. The scaling relations used by \citet{Griessmeier:2007b,Griessmeier:2011} are appropriate for sources with ages $t$\,$>$0.5 Gyr. We find that the disparity between the expected flux densities, assuming a common emission frequency, is well with our uncertainties (factors $\leq$3) for objects with ages similar to that of the Sun (2.0\,Gyr\,$<$\,$t$\,$<$\,5.0\,Gyr) than for the sources that are either very young or very old. Also note that the flux densities predicted by \citet{Reiners:2010} can be much higher than ours, even though our frequencies are in agreement, this is because they assume the stellar wind velocity to be that of the Solar wind at the location of the Earth -- this is a significant over estimate for close--in planets where the wind is still accelerating.  

\section{Comparison to observations}\label{sec:obs}

\begin{table*}
\caption{The properties of the 18 exoplanet systems expected to be detectable in the circular polarisation survey by \elenct. Included values are the right ascension and declination (RA and Dec), the mass ($M_p$) and radius ($R_p$) of the planet, the semi-major axis ($a$), the distance to the system ($d$), the age of the host star ($t$), the predicted maximum emission frequency, $\nu_c^{\text{max}}$, and flux density from our kinetic, $S_{\nu,\text{kin}}$, and magnetic, $S_{\nu,\text{mag}}$, models, as well as the best 3$\sigma$ limit available at 150\ MHz from the literature, $S_{150}$ ( Hal13: \citet{Hallinan:2013}; Sir14: \citet{Sirothia:2014}; Mur15: \citet{Murphy:2015}). The values for the full sample of sources is available online.}
\begin{tabular}{llllllrrrrrll}
\hline
Name          & RA			 & Dec	& $M_p$ 	& $R_p$     & $a$    & $d$      & $t$ & $\nu_c^{\text{max}}$& $S_{\nu,\text{kin}}$ & $S_{\nu,\text{mag}}$&  $S_{150}$ & Reference\\
              & (J2000) 	 & (J2000)		& ($M_J$)& ($R_J$) & (AU) & (pc) & (Gyr)   &  (MHz) & (mJy) & (mJy) & (mJy)    & \\
\hline
BD+20 1790 b&07:23:44&+20:24:59&07.33&1.04&0.07&25.4&0.057&1097&1.4&209.1&&\\
GJ 876 b	&22:53:13&$-$14:15:12&02.28&1.07&0.21&4.7&2.5&64&1.8&0.1&$<$17.3&Mur15\\
HATS--24 b	&17:55:34&$-$61:44:50&02.44&1.49&0.03&510&0.88&61&0.0&3.4&&\\
HD 130322 b	&14:47:33&$-$00:16:53&01.32&1.06&0.09&29.76&0.35&58&0.4&9.4&&\\
HD 210702 b	&22:11:51&+16:02:26&02.18&1.08&1.20&55.93&1.4&78&0.0&4.2&&\\
HD 5891 b	&01:00:33&+20:17:33&08.74&1.03&0.76&150.6&1.5&409&0.0&2.0&&\\
HD 73256 b	&08:36:23&$-$30:02:15&02.15&1.11&0.04&36.52&0.83&74&0.3&64.2&$<$4.0&Sir14\\
OGLE2--TR--L9 b&11:07:55&$-$61:08:47&04.50&1.61&0.03&900&0.66&132&0.0&5.0&&\\
Pr 0211 b	&08:42:11&+19:16:37&01.88&1.10&0.03&170&0.79&64&0.0&4.8&&\\
TAP 26 b	&04:18:52&+17:23:17&01.91&1.09&0.10&147&0.017&351&0.1&278.0&&\\
tau Boo b	&13:47:16&+17:27:25&04.97&1.10&0.05&15.6&2.52&159&0.2&269.2&$<$1.2&Hal13\\
V830 Tau b	&04:33:10&+24:33:43&00.70&0.99&0.06&150&0.002&198&1.6&465.0&&\\
WASP--14 b	&14:33:06&+21:53:41&07.34&1.28&0.04&160&0.75&284&0.0&10.2&&\\
WASP--140 b	&04:01:33&$-$20:27:03&02.44&1.44&0.03&180&1.6&60&0.0&4.4&&\\
WASP--167 b	&13:04:10&$-$35:32:58&08.00&1.58&0.04&381&1.29&240&0.0&10.9&&\\
WASP--18 b	&01:37:25&$-$45:40:40&10.43&1.16&0.02&105.49&0.63&504&0.1&287.1&$<$3.4&Mur15\\
WASP--43 b	&10:19:38&$-$09:48:22&01.78&0.93&0.01&80&0.4&98&0.3&173.6&&\\
WASP--77 A b	&02:28:37&$-$07:03:38&01.76&1.21&0.02&93&1.0&59&0.1&109.6&&\\
\hline
\end{tabular}\label{tab:exo_prop}
\end{table*}

In Figure (\ref{fig:tele_compare}) we focus on the exoplanets expected to produce the brightest radio emission at observable frequencies; again the values from both kinetic (\ref{fig:tele_compare}a) and magnetic (\ref{fig:tele_compare}b) RBL are shown. The 5$\sigma$ sensitivity limits of three low-frequency radio telescopes are represented by the green (GMRT), grey (Low Frequency Array; LOFAR), and blue (MWA) shaded regions.  We use the measured sensitivities reported for the MWA in \citet{Lenc:2017}, for LOFAR in \citet{vanHaarlem:2013}, and for the GMRT in \citet{Gupta:2017}, scaling each to an integration time of 8 hours. 

The integration time is chosen to maximise the sensitivity of the instruments while also taking into account the expected variability of the exoplanet emission. Due to the narrow beaming of CMI emission, it is expected that observers will only be able to detect the expected radio emission during orbital phases where the active magnetic field lines are suitably oriented relative to the line of sight to the planet. Simulating the expected radio emission properties for exoplanets, \citet{Hess:2011} showed that the detectable emission only covers a few percent of the orbital phase. The majority of the exoplanets that are expected to produce observable radio emission are Hot Jupiter systems with orbital periods of $\leq$\,7\,days \citep{Schneider:2011}. Assuming the emission is visible for 5$\%$ of the Hot Jupiter orbit, the longest time-scale we expect to see variability is $\sim$8\,hours.

The generally higher expected flux densities using the magnetic RBL means that there is a much larger number of potentially detectable planets for this case. At maximum emission frequencies $>$\,10\,MHz and using the magnetic RBL, 11 exoplanets are expected to produce $S_{\nu}$\,$>$\,100\,mJy level emission, 18 exoplanets with 10
,mJy$<$\,$S_{\nu}$\,$<$100\,mJy, and 60 are expected to produce emission at 1\,mJy$<$\,$S_{\nu}$\,$<$10\,mJy levels. In the kinetic case the maximum predicted flux density is 3.0\,mJy, with 5 objects expected to produce emission at $S_{\nu}$\,$>$1\,mJy level.  The youngest Hot Jupiters in our sample are expected to produce the highest radio powers in either the kinetic or magnetic cases. However, accounting for the distances to these sources, which can be large ($d$\,$>$100\,pc), we find that slightly older ($a$\,$\sim$2\,Gyr) Hot Jupiters that are nearby ($d$\,$<$20\,pc) are expected to produce radio emission with similar flux densities (e.g. GJ 876 b for kinetic RBL and $\tau$ Boo b for magnetic RBL). In the following we outline two observational campaigns to look for the radio emission predicted in section (\ref{sec:model}).

\subsection{MWA observations at 200 MHz}\label{sec:mwa_obs}

As demonstrated in \citet{Lynch:2017}, because the MWA is confusion limited, the sensitivity of this instrument is much greater in circular polarisation imaging than in total intensity. This makes exoplanets, whose emission is expected to be highly circularly polarised, prime candidates for MWA circular polarisations studies. \elenct recently completed the first all-sky survey in circular polarisation using archival visibility data from the Galactic and Extragalactic All-Sky MWA Survey (GLEAM, \citealt{Hurley-Walker:2017}). The survey covers declinations south of $+$30$^{\circ}$ and north of $-86^{\circ}$ with a total sky coverage of 30900 sq. degrees. \elenct used the visibility data for the 169\,--\,200\,MHz and 200\,--\,230\,MHz frequency bands, creating a final mosaic of the sky centred at 200\,MHz. This survey achieved an angular resolution of $\sim$3$^{\prime}$ and typical sensitivity of 3.0\,mJy over most of the survey region for the mosaic image. The data reduction strategy used for this survey is described in detail by \elenct.

In Figure (\ref{fig:tele_compare}) we identify the group of exoplanets expected to be detectable the survey by \elenct, assuming a 5$\sigma$ value of 15\,mJy. This selection takes into account the expected uncertainties for both the maximum emission frequency and radio flux density, selecting sources with maximum emission frequencies $\nu$\,$>$\,56\,MHz and flux densities $S_{\nu}$\,$>$\,1.5\,mJy. This sample includes 18 sources, whose physical parameters are listed in Table (\ref{tab:exo_prop}). All of these objects are Jupiter-size planets, the majority of which have masses $M_P$\,$>$\,$M_J$. Roughly $\sim$78$\%$ of the selected sources are also Hot Jupiters with semi-major axes $a$\,$<$\,0.1\,AU; all of them are young with ages $\leq$\,2.5\,Gyr. The exoplanet distance does not seem to be as restrictive as suggested by \citet{Griessmeier:2007b} when determining the brightest candidates, only 28$\%$ of our selected sources have distances $<$\,50\,pc. Our calculations show that young sources with small orbital radii can be located at distances of $\gtrsim$100\,pc and still produce detectable emission.

The predicted maximum emission frequencies and radio flux densities from the kinetic and magnetic RBL are also listed in Table (\ref{tab:exo_prop}). Only V830 Tau b and BD+20 1790 b are predicted to produce observable levels of radio emission using either the magnetic and kinetic RBL. Both V830 Tau b and BD+20 1790 b are very young and expected to encounter dense stellar winds and strong magnetic fields which inject significant energy flux into their magnetospheres. For only one source, GJ 876 b,  does the kinetic RBL predict detectable radio emission while the magnetic RBL does not. Comparing the dissipated magnetic and kinetic power as a function of distance its clear that GJ 876 b is located at a sufficient distance from its host star such that the kinetic RBL dominates over the magnetic RBL; for the 17 other sources the opposite is true.

\begin{table}
  \caption{Our measured 3$\sigma$ flux density, $S_{\nu}$, and derived luminosity, $L$,limits for both our MWA and GMRT observations.}
  \label{tab:obs_res}
  \begin{tabular}{p{0.28\columnwidth}p{0.28\columnwidth}p{0.28\columnwidth}}
  \hline
  	Name           & $S_{\nu}$ & $L$ \\
                   & (mJy)      & (erg s$^{-1}$) 		\\
    \hline
    \multicolumn{3}{r}{MWA limits at 200 MHz}\\
    \hline
     BD+20 1790 b  & <12.4      & <1.4$\times 10^{23}$\\
     GJ 876 b      & <4.5       & <1.8$\times 10^{21}$\\
     HATS--24 b     & <18.2      & <8.5$\times 10^{25}$\\  
     HD 130322 b   & <8.1       & <1.3$\times 10^{23}$ \\
     HD 210702 b   & <14.5      & <8.2$\times 10^{23}$  \\
     HD 5891 b	   & <8.2       & <3.4$\times 10^{24}$	\\   
     HD 73256 b    & <6.8       & <1.6$\times 10^{23}$	\\
     OGLE2--TR--L9 b & <13.7      & <2.0$\times 10^{26}$	\\
     Pr 0211 b     & <14.7      & <7.7$\times 10^{24}$	\\
     TAP 26 b	   & <16.8      & <6.5$\times 10^{24}$	\\
     tau Boo b     & <19.0      & <8.3$\times 10^{22}$	\\
     V830 Tau b    & <45.3      & <1.8$\times 10^{25}$	\\
     WASP--14 b     & <32.0      & <1.5$\times 10^{25}$\\
     WASP--140 b    & <4.7       & <2.7$\times 10^{24}$	\\
     WASP--167 b    & <7.1       & <1.9$\times 10^{25}$\\
     WASP--18 b     & <4.1       & <8.2$\times 10^{23}$	\\
     WASP--43 b     & <6.8       & <7.8$\times 10^{23}$  \\
     WASP--77 A b   & <6.0       & <9.2$\times 10^{23}$	\\
     \hline
     \multicolumn{3}{r}{GMRT limits at 150 MHz}\\
     \hline
     V830 Tau b    & <4.5      & <5.9$\times 10^{24}$	\\
     \hline
     
  \end{tabular}
 \end{table}

Only 4 of these sources have previous radio upper limits, also listed in (\ref{tab:exo_prop}), and so the limits reported here are new for most of these sources. Using the mosaic images from \elenct, we did a targeted search of the 18 sources for significant circularly polarised emission above the estimated local image noise. No radio source was detected at the location of the exoplanets we considered. To place limits on the circular polarised emission from these sources, we measured the rms in a box centred on the location of the source and having a size several times the synthesised beam. The 3$\sigma$ upper limits are listed in Table (\ref{tab:obs_res}). We also list in Table (\ref{tab:obs_res}) the luminosity limits assuming the emission is steady and we ignore any beaming effects ($\Omega$\,=\,4$\pi$). For 10 out of the 18 sources we targeted, the upper limits are lower than the expected flux densities from either the magnetic or kinetic RBL. Taking into account the uncertainty associated with the flux density estimates, if the predicted emission levels are an order-of-magnitude lower only 7 of those sources still have emission that is above our 3$\sigma$ limits. 

\begin{figure*}
\centering
	\includegraphics[width=0.6\linewidth]{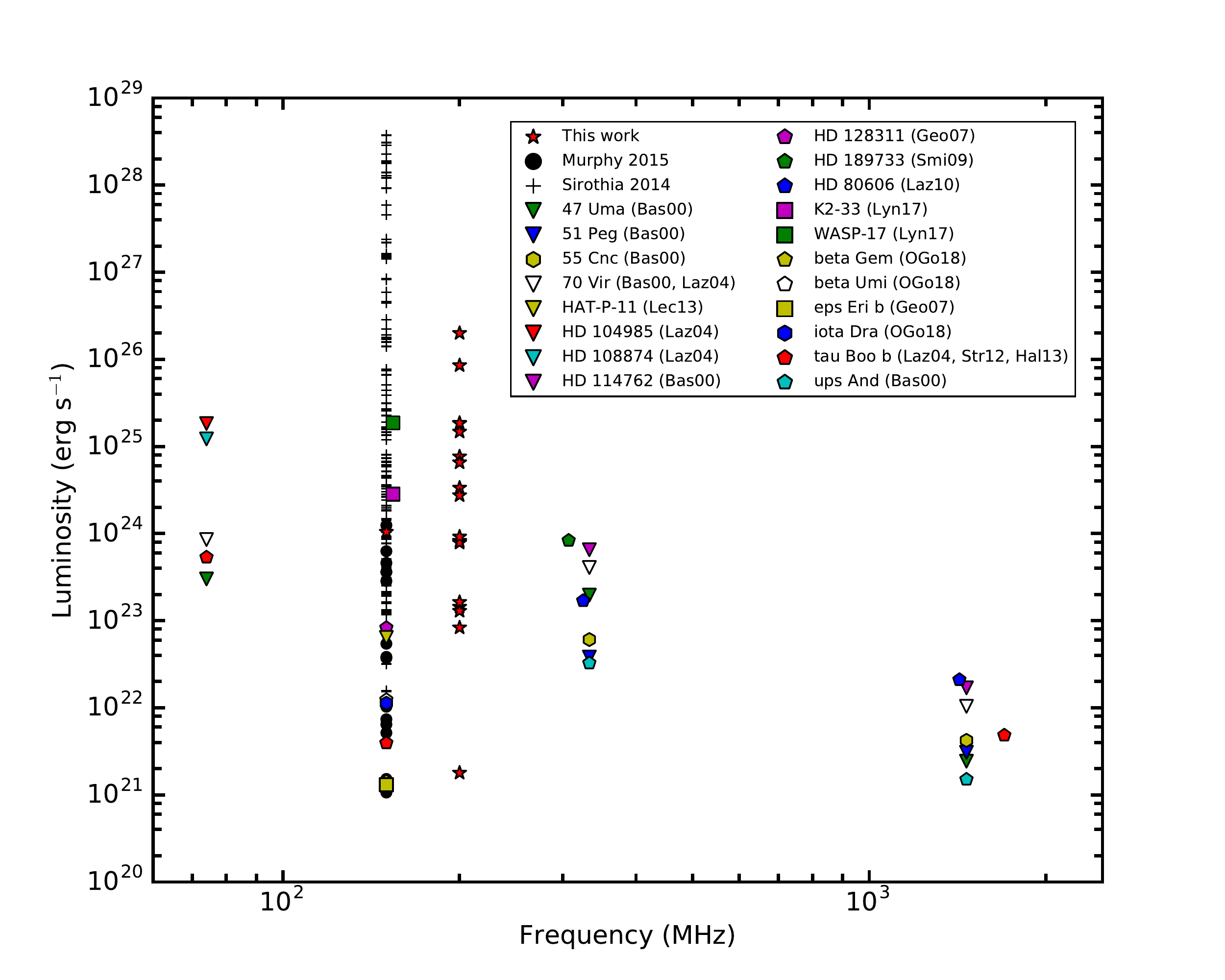}
    \caption{Limits on the radio luminosity from exoplanets; the limits presented in this paper are given by the red stars. The references in the legend are: Bas00 \citep{Bastian:2000}; Laz04 \citep{Lazio:2004}; Geo07 \citep{George:2007}; Smi09 \citep{Smith:2009}; Laz10 \citep{Lazio:2010}; Str12 \citep{Stroe:2012}; Hal13 \citep{Hallinan:2013}; Lec13 \citep{Lecavelier:2013}; Lyn17 \citep{Lynch:2017}; OGo18 \citep{OGorman:2018}.} \label{fig:lum_limits}
\end{figure*}

In Figure (\ref{fig:lum_limits}) we compare the luminosity limits from the literature to those reported here. The MWA upper limits are unique in that they place limits on the exoplanet radio emission within a new frequency range (169\,--\,230\,MHz). This is important because of the expected sharp cutoff in emission frequency. Complete coverage of all available low-frequency observing bands is necessary to rule out non-detections due to observing at the wrong frequency. The MWA limits we report here are comparable to the best limits already reported in the literature.

\subsection{GMRT observations of V830 Tau b}

The location of V830 Tau b in the sky is such that the GLEAM survey only observed its position during a single epoch with a total time on source of 12 minutes. The lack of coverage for this position contributes greatly to its high rms measured in the circular polarisation survey of \elenct. As this is an interesting source, given its high predicted radio flux densities, we observed this source using the GMRT to try to place better constraints on the radio emission from this source.

 We observed V830 Tau b in three 8-hour observing epochs on Dec 03, 05, 06 2017 for a total of on-source time of 24 hours. Due to the highly beamed nature of the radio emission we expect that the emission will only be observable over orbital and/or rotational phases where the beam of emission is optimally aligned with our line of sight. If V830 Tau b is tidally locked, we expect the rotational period to be equivalent to its 4.93 day period. Therefore we asked for our observations to be scheduled in such a way to maximise our orbital coverage. With the 24 hours of observational time we were granted, we achieved coverage of 20$\%$ of the orbit. For each of our observations we observed over a bandwidth of 16 MHz centred at 156 MHz. We observed the primary calibrators 3C48, 3C147 and a secondary calibrator PKS 0428+205. To reduce and image the data we used the Source Peeling and Atmospheric Modelling (SPAM) software \citep{Intema:2009}. SPAM is a data reduction python module that uses the NRAO Astronomical Image Processing Software \citep[AIPS;][]{Greisen:2003} via the ParselTongue interface \citep{Kettenis:2006} and Obit \citep{Cotton:2008}.

We give a brief summary of the data reduction process of the SPAM pipeline; a more detailed outline of the pipeline is given by \citet{Intema:2017}. The SPAM reduction pipeline calibrates and images the data in a two--step process. During the first step, calibration solutions are derived for our two primary calibrator sources, 3C47 and 3C148. These solutions are found incrementally over iterations of RFI flagging and derivations of the complex gain and bandpass solutions. SPAM sets the flux scale by assuming a point source model and uses the flux densities and spectral indices from \citet{Scaife:2012}. To select the best calibrator scan a weight factor was calculated from the number of active antennas and the inverse variance of the gain normalised amplitudes. The best calibrator scan is the one with the highest weight factor.  The calibration tables from the selected primary calibrator were then applied to the target field scans and basic RFI flagging was applied. The pipeline then performs initial phase calibration and astrometry correction of the target field using a sky model derived from the NVSS \citep{Condon:1998}.  

The second step begins with three rounds of direction-independent phase-only self-calibration, during which various automated flagging routines are used between the cycles of imaging and self-calibration to reduce the residual RFI and clip statistical outliers. SPAM then performs facet-based direction-dependent calibration on strong sources identified within the primary beam FWHM. The solutions from this calibration are used to fit to a global ionospheric model.  Using this model, a set of gain corrections is generated for the image facets. These gain corrections are applied during a final round of imaging and de-convolution of the target field.

We calibrated each observing epoch independently using SPAM and then median-stacked the final total intensity images to produce a deep image. We achieved an rms of $\sim$2.0\,mJy in the individual, 8\,hour images, and an rms of 1.5\,mJy in the deep 24-hour image. We did not detect any significant emission at the location of the planet to a 3$\sigma$ level in any of the these images; the 3$\sigma$ limit from our deep image and associated luminosity limit are reported in Table (\ref{tab:obs_res}). Our new limit is about an order of magnitude better than the limit placed by the MWA observations in section (\ref{sec:mwa_obs}). With these more sensitive observations we can more confidently say that if V830 Tau b had produced radio emission beamed into our line of sight at the levels predicted by our models, including their uncertainty, we would have made a significant detection ($>$10$\sigma$) of this emission.

\section{Conclusions}
We presented predictions for the radio emission properties for the currently known population of exoplanets orbiting dwarf type stars. We report the predicted maximum emission frequencies and the associated radio flux densities for this population of sources. Similar to previous modelling efforts of the comprehensive population of known exoplanets, we find that the expected radio flux density decreases with the associated maximum emission frequency. This trend highlights a deficient number of Jupiter or larger sized planets orbiting their host stars at small radii. For most objects the magnetic RBL predicts much brighter radio emission than the kinetic RBL, however in a few cases the exoplanet is located at a sufficient distance away from its host star so the kinetic power dissipated dominates the star--planet interaction. The brightest emission is predicted for Hot Jupiters orbiting young ($t$\,$\leq$2.5\,Gyr) stars. 

We also report upper limits from two observational efforts to detect radio emission from exoplanets. Using a recent all-sky circular polarisation survey from the MWA, we place the first upper limits on 169\,--\,230\,MHz emission from exoplanets. These limits are comparable to previous limits in the literature at other frequencies. We also carried out a targeted observational campaign of V830 Tau b using the GMRT. This source is of particular interest because of the predicted bright radio emission. We place the first low-frequency radio limit on emission from this source.

Not much can be gleamed from these upper limits, as the uncertainties in the predicted radio flux densities are large, and many other factors could explain the non-detections. The large uncertainty associated with the maximum emission frequency makes it such that the range of possible emission frequencies is larger than the bandwidth of current instruments. Better estimates of the exoplanet masses could alleviate this problem. Additionally, the scaling relation of \citet{Christensen:2009} appears consistent with magnetic field measurement from Solar System planets to fully convective stars, yet it is unconstrained over the mass gap occupied by massive exoplanets and rapidly rotating brown dwarfs. CMI emission has now been detected from the coolest brown dwarf objects (L and T dwarfs), providing measurements for magnetic field strengths of objects with masses between that of Jupiter and low mass stars \citep{Kao:2016}. These magnetic field strengths tentatively suggest a deviation from the scaling relation of \citet{Christensen:2009}. Future radio studies of such objects could place better constraints on this scaling relation and allow for better predictions of the maximum emission frequencies for Hot Jupiters. Further, much could be gained by pushing down limits at the lowest observable frequencies since we know Jupiter can produce emission at these frequencies \citep{Zarka:2007} and eliminates the uncertainty as to whether planets can produce higher frequency emission. 

For the majority of exoplanets in our sample, the magnetic RBL predicts much brighter radio emission than the kinetic RBL. However, as noted in the literature, saturation of magnetospheric convection in Hot Jupiter atmospheres could cause these systems to be unable to dissipate the total incident magnetic energy from the stellar wind \citep{Jardine:2008, Nichols:2016}. If this is true, than the magnetic RBL significantly overestimates the expected flux densities from these systems; \citet{Nichols:2016} predicts flux densities on the order of 1\,mJy in the case of convection saturation. Even worse still, \citet{Weber:2017} suggests that these massive planets will have extended ionospheres with plasma densities large enough to prevent the propagation of CMI emission. Counter to that point, however,
\citet{Yadav:2017} note that the luminosity relations used to calculate the magnetic field strengths for Hot Jupiter systems do not take into account that most of these objects are inflated and could have much stronger magnetic field strengths. This might alleviate the issue presented by \citet{Weber:2017}. 

Lastly, the beaming of the emission could also explain the non-detections. Only a small percentage of the exoplanet orbits were covered by either the GMRT or MWA observations reported here. Given this, we cannot rule out that there is an optimal orbital phase within which we would have detected the expected emission. Observations that cover the full orbit of young Hot Jupiter systems would be able to rule out this scenario.

The increased sensitivity expected through future upgrades to current low-frequency telescopes and ultimately the Square Kilometre Array \citep{Lazio:2009} will extend sensitivity limits into a flux density and frequency regime where many more exoplanets are predicted to produce observable levels of radio emission. The greater number of potentially observable exoplanets will help to alleviate some of the issues associated with these non-detection and increases the likelihood of a radio detection.

\section*{Acknowledgements}

This scientific work makes use of the Murchison Radio-astronomy Observatory, operated by CSIRO. We acknowledge the Wajarri Yamatji people as the traditional owners of the Observatory site. Support for the operation of the MWA is provided by the Australian Government (NCRIS), under a contract to Curtin University administered by Astronomy Australia Limited. We acknowledge the Pawsey Supercomputing Centre which is supported by the Western Australian and Australian Governments. This research was conducted by the Australian Research Council Centre of Excellence for All-sky Astrophysics (CAASTRO), through project number CE110001020. We thank the staff of the GMRT that made these observations possible. GMRT is run by the National Centre for Radio Astrophysics of the Tata Institute of Fundamental Research. DK was supported by NSF grant AST-1412421. T.M. acknowledges the support of the Australian Research Council through grant FT150100099. This research has made use of the NASA Exoplanet Archive, which is operated by the California Institute of Technology, under contract with the National Aeronautics and Space Administration under the Exoplanet Exploration Program. The authors thank the anonymous referee for providing helpful comments on the original version of this paper.




\bibliographystyle{mnras}
\newcommand{\noop}[1]{}


\bsp	
\label{lastpage}
\end{document}